\documentclass [12 pt]{article}
\def\be{\begin{equation}}
\def\eea{\end{eqnarray}}
\def\bea{\begin{eqnarray}}
\def\ee{\end{equation}}

\def\iota{\imath}
\usepackage[psamsfonts]{amssymb}
\usepackage{amsmath}
\author{M. Amooshahi$^{1}$
\footnote{amooshahi@sci.ui.ac.ir} , B. Nasr Esfahani $^{1}$
\footnote{ba$\_$nasre@sci.ui.ac.ir}\\
{\small $^{1}$Department of Physics, Faculty of Science, University
of Isfahan,}\\{\small Hezar Jarib Ave., Isfahan, Iran}}
\title{Canonical quantization of the electromagnetic field in the presence of non-dispersive bi-anisotropic inhomogeneous  magnetodielectric media}
\begin{document}
\maketitle
\begin{abstract}

By introducing  a suitable Lagrangian, a canonical quantization of
the electromagnetic field in the presence of a non-dispersive
bi-anisotropic inhomogeneous magnetodielectric medium is
investigated. A tensor  projection operator is defined  and the
commutation relation between the vector potential and its
canonically conjugate variable is written in terms of the projection
operator. The quantization method is generalized in the presence of
the atomic systems. The spontaneous emission of a two-level atom
located in a non-dispersive anisotropic megnetodielectric medium is studied.
\end{abstract}
\section{introduction}

The quantum properties of the electromagnetic field apparently are
influenced by the presence of magnetodielectric media. For examples
the spontaneous decay constant of atomic systems\cite{1}-\cite{5},
the Casimir effect \cite{6} and the statistical properties of light
are modified in the presence of a polarizable or magnetizable medium
\cite{7}-\cite{9}. The quantization of electromagnetic field is
considered usually in the presence of two types of media,  the
dispersive lossy magnetodielectric media and non-dispersive one. In
dispersive magnetodielectric media there is a temporally nonlocal
relationship between the polarization (magnetization ) field  and
the electric (magnetic) field  \cite{10}-\cite{13}. In this type of
media, in order to  inclusion the lossy effects, the medium is
modeled by a collection of harmonic oscillators and both the
electromagnetic field and the responsive medium are quantized,
\cite{14}-\cite{20}. These quantization approaches cover both
isotropic and inhomogeneous anisotropic media. In these schemes the
permittivity and permeability of the medium is calculated in terms
of the classical parameters    applied in the lagrangian or
Hamiltonian of
the total system.\\

In  the dispersionless dielectric  media the relationship between
the polarization field and the electric field is temporally local.
To quantize the electromagnetic field in the presence of such media,
the medium is not quantized directly. The polarization  effects of
the medium is introduced, in the lagrangian density of the total
system, only by the its linear and nonlinear susceptibility tensors
\cite{21}-\cite{28}. In this case the Euler-Lagrange equations are
the macroscopic Maxwell equations in the presence of the
non-dispersive medium. This method has been developed for  linear or
non-linear isotropic dielectric media. In the dispersionless
responsive media there is a class of linear polarizable and
magnetizable media that the electric polarization (magnetic
polarization) is related linearly to both the electric and magnetic
fields. Generally, the electromagnetic properties of these media are
described by four tensors of the second rank. This kind of media are
known as bi-anisotropic magnetodielectric media. In this paper
proposing a Lagrangian density a fully canonical quantization of
electromagnetic field is represented in the presence of such media.
It is worthy to point out that the form of Maxwell's equations in
the curved space-time , written in the cartesian coordinates,  in
the absence of any responsive medium, are identical to the form of
Maxwell's equations in flat space-time  in the presence of a
bi-anisotropic magnetodielectric media\cite{29,30}. Therefore the
quantization method demonstrated  in the present work is applicable
for the quantized electromagnetic field in the curved
space-time  in the absence of any responsive medium. The organization of this paper is as follows:\\

In  sec.2  a Lagrangian for the electromagnetic field in the
presence of a bi-anisotropic magnetodielectric medium is proposed
and a classical treatment a bout the problem is achieved. In sec.3
using the lagrangian introduced in section 2 a canonical
quantization of the electromagnetic field in the presence of a
bi-anisotropic inhomogeneous magnetodielectric medium is
represented. In sec. 4 the quantization method is generalized in the
presence of the atomic systems. In sec.5  the spontaneous emission
of an initially excited two-level atom located in a bi-anisotropic
medium is investigated. Finally the paper is closed by a summary in
sec.6.
\section{Classical Lagrange equations}

In the non-dispersive linear magnetodielectric media  the
displacement field ${\bf D}$ and the magnetic field
 ${\bf H}$ are related to the electric field ${\bf E}$ and
the magnetic induction field ${\bf B}$ as \cite{31}
\begin{eqnarray}\label{a.1}
      &&{\bf D}({\bf r},t)=\varepsilon^{(1)}({\bf r}){\bf E}({\bf
        r},t)+\varepsilon^{(2)}({\bf r}){\bf B}({\bf r},t),\nonumber\\
        &&{\bf H}({\bf r},t)=\mu^{(1)}({\bf r}){\bf E}({\bf
      r},t)+\mu^{(2)}({\bf r}){\bf B}({\bf r},t).
\end{eqnarray}
The electric  (the magnetic) properties of the medium are described
by the permittivities   $\varepsilon^{(1)}({\bf r}) ,
\varepsilon^{(2)}({\bf r})$  ( the permeabilities $\mu^{(1)}({\bf
r}) ,  \mu^{(2)}({\bf r})$), respectively. Generally, for
bi-anisotropic media the permittivities and permeabilities may be
appeared as four tensors of the second rank. There are some symmetry
relations between  the permittivities and permeabilities of a
medium. The most important symmetry relations between the
permittivities and the permeabilities of a non-dispersive
magnetodielectric medium  are the Onsager's relations \cite{31}
\begin{equation}\label{a.2}
       \varepsilon^{(1)}_{ij}=\varepsilon^{(1)}_{ji}\hspace{1.50
         cm}\mu^{(2)}_{ij}=\mu^{(2)}_{ji} \hspace{1.50
       cm}\varepsilon^{(2)}_{ij}=-\mu^{(1)}_{ji}
\end{equation}
A realization of the constitutive relations (\ref{a.1}) and the
symmetry relations (\ref{a.2}) is the electromagnetic field in a
curved space-time  in the absence of any responsive medium. It is
well known that  Maxwell's equations in a curved space-time in the
absence of external sources, when  the cartesian coordinates is
used, are as the form \cite{29,30}
\begin{eqnarray}\label{a.3}
      &&\nabla\cdot{\bf B}=0\hspace{2.00 cm}\nabla\cdot{\bf D}=0\nonumber\\
        &&\nabla\times{\bf H}=\frac{\partial{\bf D}}{\partial t}\hspace{2.00
      cm}\nabla\times{\bf E}=-\frac{\partial{\bf B}}{\partial t}
\end{eqnarray}
where the cartesian components of the fields ${\bf D}$ and ${\bf
H}$ are defined by
\begin{eqnarray}\label{a.4}
      D_i=\varepsilon_{ij}E_j+({\bf G}\times {\bf H})_i\hspace{2.00
      cm}B_i=\mu_{ij}H_j+({\bf G}\times {\bf E})_i
\end{eqnarray}
with
\begin{eqnarray}\label{a.5}
      &&\varepsilon_{ij}=\mu_{ij}=-\sqrt{-g}\
        \frac{g^{ij}}{g_{00}}\hspace{1.50 cm}i,j=1,2,3\nonumber\\
      &&G_i= -\frac{g_{0i}}{g_{00}}\hspace{3.50 cm}i=1,2,3
\end{eqnarray}
 where $g_{\mu\nu}$  is the space-time metric and $g$ is its determinant. The definitions
 (\ref{a.4}) for the fields {\bf D} and {\bf H} can be rewritten in the form of
 constitutive relations (\ref{a.1}) with the
 tensors $ \varepsilon^{(1)} ,\ \varepsilon^{(2)}, \mu^{(1)},\
 \mu^{(2)}$  given by
\begin{eqnarray}\label{a.6}
       && \varepsilon^{(1)}_{ij}=-\sqrt{-g}\
        \frac{g^{ij}}{g_{00}}-\epsilon_{i\alpha\beta}\epsilon_{mnj}\
         \frac{g_{0\alpha}\  g_{0n}\ g_{\beta m}}{g_{00}\sqrt{-g}}\nonumber\\
          &&\varepsilon^{(2)}_{ij}=-\epsilon_{imn}\frac{g_{nj}\
          g_{0m}}{\sqrt{-g}}\nonumber\\
         &&\mu^{(1)}_{ij}=-\epsilon_{mnj}\ \frac{g_{0n}\
        g_{im}}{\sqrt{-g}}\nonumber\\
       &&\mu^{(2)}_{ij}=-\frac{g_{00}}{\sqrt{-g}}\ g_{ij}
\end{eqnarray}
where $\epsilon_{ijk}$ is the 3-dimensional Levi-Civita symbol. It
is easy to investigate that the tensors defined by (\ref{a.6}) obey
the Onsager's relation (\ref{a.2}). Therefore, applying  the
cartesian coordinates,  the form of Maxwell's equations in the
curved space-time  in the absence of a responsive medium is similar
to the form of  Maxwell's equations in the flat space-time  in the
presence  of a bi-anisotropic magnetodielectric medium. The above
discussions  are sufficient motivation   to investigate the
quantization of the electromagnetic field in the presence of a
bi-anisotropic magnetodielectric medium. \\

 Using the constitutive relations
(\ref{a.1}) Maxwell's equations (\ref{a.3})  can be rewritten  as
\begin{equation}\label{a.7}
      \nabla\cdot{\bf B}=0
\end{equation}
\begin{equation}\label{a.8}
      \nabla\times{\bf E}=-\frac{\partial{\bf B}}{\partial t}
\end{equation}
\begin{equation}\label{a.9}
      \nabla\cdot[\varepsilon^{(1)}({\bf r}){\bf E}({\bf
      r},t)+\varepsilon^{(2)}({\bf r}){\bf B}({\bf r},t)]=0
\end{equation}
\begin{equation}\label{a.95}
      \nabla\times[\mu^{(1)}({\bf r}){\bf
        E}({\bf r},t)+\mu^{(2)}({\bf r}){\bf B}({\bf
        r},t)]=\frac{\partial}{\partial t}[\varepsilon^{(1)}({\bf r}){\bf
      E}({\bf r},t)+\varepsilon^{(2)}({\bf r}){\bf B}({\bf r},t)]
\end{equation}
The classical Maxwell's equations (\ref{a.9}) and (\ref{a.95}) can
be obtained as a consequence of the principle of Hamilton's least
action using the Lagrangian
\begin{eqnarray}\label{a1}
      &&L(t)= \int d^3r\pounds(\textbf{r},t)=\int d^3r \{
       \frac{1}{2}\varepsilon^{(1)}_{ij}(\textbf{r})
        E^i(\textbf{r},t)E^j(\textbf{r},t)\\\nonumber
        &&+\frac{1}{2}\varepsilon^{(2)}_{ij}(\textbf{r})
         E^i(\textbf{r},t)B^j(\textbf{r},t)-\frac{1}{2}\mu^{(1)}_{ij}(\textbf{r})B^i(\textbf{r},t)
       E^j(\textbf{r},t)-\frac{1}{2}\mu^{(2)}_{ij}(\textbf{r})
      B^i(\textbf{r},t)B^j(\textbf{r},t)\}.
\end{eqnarray}
where the Einstein sum rule has been applied. In (\ref{a1}),
$\textbf{E}=-\frac{\partial\textbf{A}}{\partial t}-\nabla\varphi$
and $ \textbf{B}=\nabla\times\textbf{A}$.  This means that the
vector potential ${\bf A}$ and the scalar potential $\varphi$
constitute the degrees of freedom of the electromagnetic field. It
is easy to show that the Lagrange equation for the  variable
$\varphi$  leads to the Gauss' law (\ref{a.9}) while, using the
Onsager's relations (\ref{a.2}), the Lagrange equation for the
vector potential ${\bf A}$ gives the Maxwell's equation
(\ref{a.95}). \\

As it is clear in the present formalism the non-dispersive
magnetodielectric medium is not quantized directly and the effect of
the medium is appeared as the classical tensors
$\varepsilon^{(1)},\varepsilon^{(2)}, \mu^{(1)}$ and $\mu^{(2)}$.
However in the quantization of the electromagnetic field in the
presence of dispersive lossy media, the medium itself is quantized
\cite{14}-\cite{20}.
\subsection{Gauge fixing}
For a consistent canonical quantization of the electromagnetic
field, we need the extra degrees of freedom to be eliminated from
the above Lagrangian  using some appropriate gauge conditions. Here
we apply the gauge condition
\begin{equation}\label{a4}
      \nabla\cdot\left(
      \varepsilon^{(1)}(\textbf{r})\textbf{A}(\textbf{r},t)\right)=0,
\end{equation}
where previously has been used by Glauber and et al \cite{25}.
Combination of this gauge with the Gauss' law (\ref{a.9}) leads to
expression
\begin{equation}\label{a5}
      \varphi(\textbf{r},t)=-\int d^3r'\ G(\textbf{r}, \textbf{r}')\
      \nabla'\cdot\left[ \varepsilon^{(2)}(\textbf{r}')\ \nabla'\times
      \textbf{A}(\textbf{r}',t)\right],
\end{equation}
for the scalar potential, where $G(\textbf{r}, \textbf{r}')$ is the
Green function satisfying the differential  equation
\begin{equation}\label{a6}
      \nabla\cdot\left(\varepsilon^{(1)}(\textbf{r})\ \nabla
      G(\textbf{r},\textbf{r}')\right)=-\delta(\textbf{r}-\textbf{r}').
\end{equation}
Let us investigate the symmetry property
\begin{equation}\label{a7}
      G(\textbf{r},\textbf{r}')=G(\textbf{r}',\textbf{r})
\end{equation}
for  this Green function. To investigate  this symmetry feature we
consider a generalized Green's theorem for two arbitrary scalar
functions $\varphi_1({\bf q})$ and $\varphi_2({\bf q})$ as
\begin{eqnarray}\label{a8}
      &&\int_V d^3q \left (\varphi_1\ \nabla\cdot
       (\varepsilon^{(1)}\vec{\nabla}\varphi_2)-\varphi_2\ \nabla\cdot
        (\varepsilon^{(1)}\vec{\nabla}\varphi_1)\right)\nonumber\\
       &&=\oint_s\left( \varphi_1\ \varepsilon^{(1)}
      \vec{\nabla}\varphi_2-\varphi_2\ \varepsilon^{(1)}
     \vec{\nabla}\varphi_1\right)\cdot \hat{n}\ d s
\end{eqnarray}
which can be proved using the symmetry property of the tensor
$\varepsilon^{(1)}$, that is
$\varepsilon^{(1)}_{ij}=\varepsilon^{(1)}_{ji}$.
 Choosing $ \varphi_1(\textbf{q})=G(\textbf{q},\textbf{r})$ and
$\varphi_2(\textbf{q})=G(\textbf{q},\textbf{r}')$, and regarding
(\ref{a6}) and the boundary condition that the Green function should
vanish on a surface located at infinity, the expected symmetry
property (\ref{a7}) is deduced.

Let us expediently resolve  a squared integrable vector field
$\textbf{F}(\textbf{r})$  into two components, as $\textbf{F}=
\textbf{F}^\|+ \textbf{F}^\bot$. The component $\textbf{F}^\|$ is
defined by
\begin{equation}\label{a9}
      \textbf {F}^\|(\textbf{r})=-\int d^3r'\nabla G(\textbf{r},\textbf{r}')
        \nabla'\cdot\ \left(\varepsilon^{(1)}(\textbf{r}')\
      \textbf{F}(\textbf{r}')\right)
\end{equation}
which is a conservative field. The other part $\textbf{F}^\bot$ is
given by
\begin{equation}\label{a10}
       \textbf{F}^\bot=\textbf{F}-\textbf{F}^\|
\end{equation}
which satisfies the relation $
\nabla\cdot\left(\varepsilon^{(1)}(\textbf{r})\
\textbf{F}^\bot(\textbf{r})\right)=0$. we call $ \textbf{F}^\|$ and
$ \textbf{F}^\bot$ as  $\varepsilon^{(1)}$- longitudinal  and
$\varepsilon^{(1)}$- transverse components of the vector field
$\textbf{F}$, respectively. Conveniently,  the tensor projection
operators $ P^\bot_{ij}(\textbf{r},\textbf{r}')$ and
$P^\|_{ij}(\textbf{r},\textbf{r}')$ are introduced as
\begin{eqnarray}\label{a11}
       P^\|_{ij}(\textbf{r},\textbf{r}')&=&\sum_{l=1}^3\left[
        \varepsilon^{(1)}_{lj}(\textbf{r}')
          \frac{\partial}{\partial x_i}\frac{\partial}{x'_l}
          G(\textbf{r},\textbf{r}')\right]\nonumber\\
        P^\bot_{ij}(\textbf{r},\textbf{r}')&=& \delta_{ij}
      \delta(\textbf{r}-\textbf{r}')-P^\|_{ij}(\textbf{r},\textbf{r}').
\end{eqnarray}
In terms of these projection operators, one can rewrite the
definitions (\ref{a9}) and (\ref{a10}) as
\begin{eqnarray}\label{a12}
       F^\|_i(\textbf{r})&=& \sum_{j=1}^3\int d^3r'\
         P^\|_{ij}(\textbf{r},\textbf{r}')\
          F_j(\textbf{r}')\hspace{1.50 cm} i=1,2,3
\end{eqnarray}
\begin{eqnarray}\label{a12.1}
         F^\bot_i(\textbf{r})&=& \sum_{j=1}^3 \int d^3r'\ P^\bot_{ij}(\textbf{r},\textbf{r}')\
       F_j(\textbf{r}')\hspace{1.50 cm} i=1,2,3.
\end{eqnarray}
Using the poisson's equation ( \ref{a6}) and the symmetry property
(\ref{a7}) it is easy to show that the projection operators $
P^\bot$ and $P^\|$ satisfy the transversality  relations
\begin{equation}\label{a13}
      \sum_{i,m=1}^3\ \frac{\partial}{\partial x_i} \left(
        \varepsilon^{(1)}_{im}(\textbf{r})\
      P^\bot_{mj}(\textbf{r},\textbf{r}')\right)=0\hspace{1.50 cm} j=1,2,3,
\end{equation}
and
\begin{equation}\label{a14}
      \sum_{j=1}^3\ \frac{\partial}{\partial x'_j}\
      P^\bot_{ij}(\textbf{r},\textbf{r}')=0\hspace{1.50 cm} i=1,2,3.
\end{equation}
The gauge condition (\ref{a4}) shows that  $- \nabla\varphi$ and $ -
\frac{\partial\textbf{A}}{\partial t}$  are  $\varepsilon^{(1)}$-
longitudinal and $\varepsilon^{(1)}$- transverse components of the
the electric field  $\textbf{E}$, respectively. This can be seen by
operating  $P^\bot$ and $P^\|$ on the electric field, that is
\begin{eqnarray}\label{a16}
      - \frac{\partial A_i}{\partial t}&=&  \sum_{j=1}^3\int d^3r'\ \
      P^\bot_{ij}(\textbf{r},\textbf{r}')\ E_j(\textbf{r}')\\
      - \frac{\partial \varphi}{ \partial x_i}&=&\sum_{j=1}^3\int d^3r'\ \
      P^\|_{ij}(\textbf{r},\textbf{r}')\ E_j(\textbf{r}')
\end{eqnarray}
According to the Eq. (\ref{a14}), one can obtain the transverse
component of any squared integrable vector field
$\textbf{F}(\textbf{r})$ as
\begin{equation}\label{a15}
       F^\top_i(\textbf{r})= \sum_{j=1}^3\int d^3r'\ \
       P^\bot_{ji}(\textbf{r},\textbf{r}')\ F_j(\textbf{r}')\hspace{1.50 cm}
       i=1,2,3.
\end{equation}
which clearly satisfies $ \nabla\cdot \textbf{F}^\top=0$.

\section{Canonical quantization}

Before beginning a canonical quantization scheme, the extra degrees
of freedom should be eliminated from the Lagrangian of the system
applying the constraints (\ref{a4}) and (\ref{a5}). This can be done
by substituting the scalar potential $\varphi$ from (\ref{a5}) into
the Lagrangian (\ref{a1}) and doing some integration by parts and
using the antisymmetry  relation
$\varepsilon^{(1)}_{ij}=-\mu^{(1)}_{ji}$. Then we have
\begin{eqnarray}\label{a18}
      &&L(t)=\int d^3r\left\{ \varepsilon^{(1)}_{ij}(\textbf{r})
        \frac{\partial A_i}{\partial t} \frac{\partial A_j}{\partial
         t}-\varepsilon^{(1)}_{ij}(\textbf{r}) \frac{\partial A_i}{\partial t}
           \left[(\varepsilon^{(1)}(\textbf{r}))^{-1}\
            \varepsilon^{(2)}(\textbf{r})\
             \nabla\times \textbf{A}(\textbf{r},t)\right]^\bot_j\right.\nonumber\\
             &&\left.- \frac{1}{2} \mu^{(2)}_{ij}( \nabla\times \textbf{A})_i (
             \nabla\times \textbf{A})_j\right\}\\
            &&-\frac{1}{2}\int d^3r \int d^3r'\ G(\textbf{r},\textbf{r}')\
           \nabla\cdot[ \varepsilon^{(2)}(\vec{r}) \nabla\times
          \textbf{A}(\textbf{r},t)]\ \nabla'\cdot\ [\varepsilon^{(2)}(\textbf{r}')
         \nabla'\times
        \textbf{A}(\textbf{r}',t)],\nonumber
\end{eqnarray}
 Having the Lagrangian density
of the system, denoted by $\pounds$, as the integrand in
(\ref{a18}), the $i$'th cartesian component of the canonical
conjugate of the dynamical variable $\textbf{A}$ can easily be
calculated in a standard way as
\begin{eqnarray}\label{a19}
      \Pi_i(\textbf{r},t)&=&\frac{\partial\pounds}{\partial
        A_i}\\
          &=&\varepsilon^{(1)}(\textbf{r})\frac{\partial A_i(\textbf{r},t)}{\partial
         t}-\varepsilon^{(1)}(\textbf{r})\left[
        (\varepsilon^{(1)}(\textbf{r}))^{-1}\ \varepsilon^{(2)}(\textbf{r})
       \nabla\times\textbf{A}(\textbf{r},t)\right]_i^\bot\nonumber
\end{eqnarray}
As it is seen, the canonical conjugate variable ${\bf \Pi}$ is a
purely transverse vector field , that is $ \nabla\cdot{\bf \Pi}=0$ ,
while according to the gauge condition (\ref{a4}) the vector
potential $\textbf{A}$ is a purely $ \varepsilon^{(1)}$-transverse
vector field. Now the canonical quantization can be accomplished in
a standard fashion by imposing the following equal time commutation
relations between the cartesian components of the conjugate
variables $ {\bf \Pi}$ and ${ \bf A}$ as
\begin{equation}\label{a20}
      [A_i({\bf r},t)\ \ \Pi_j({\bf r'},t)]= P^\bot_{ij}({\bf r},{\bf
      r'}).
\end{equation}
Because of the transversality relations (\ref{a13}) and (\ref{a14}),
it is clear that the commutation relations (\ref{a20}) is compatible
to  the conditions $ \nabla\cdot {\bf \Pi}=0$ and $ \nabla\cdot(
\varepsilon^{(1)}({\bf r}) {\bf A})=0$.  The time evolution of any
dynamical operator related to the electromagnetic field, in the
Heisenberg picture, can be obtained by using of the Hamiltonian of
the system. The Hamiltonian of the electromagnetic field can be
written in terms of $ {\bf A}$ and $ {\bf \Pi}$ as ( see the
Appendix A)
\begin{eqnarray}\label{a21}
       &&H(t)=\frac{1}{2} \int d^3r\left\{ [ \varepsilon^{(1)}({\bf
        r})]^{-1}_{ij}\ \Pi_i({\bf r},t)\ \Pi_j({\bf r},t)+
         \mu^{(2)}_{ij}({\bf r})\ ( \nabla\times \textbf{A})_i\ ( \nabla\times
          \textbf{A})_j\right\}\nonumber\\
           &&+\frac{1}{2}\int d^3r\   [ \varepsilon^{(1)}({\bf
            r})]^{-1}_{ij}\left\{\Pi_i({\bf r},t) \left[ \varepsilon^{(2)} ({\bf
            r})\nabla\times \textbf{A}\right]_j+ \left[ \varepsilon^{(2)} ({\bf r
           })\nabla\times \textbf{A}\right]_i\Pi_j({\bf r},t)\right\}\nonumber\\
          && +\frac{1}{2}  \int d^3r \  [ \varepsilon^{(1)}({\bf
        r})]^{-1}_{ij} \ [\varepsilon^{(2)} ({\bf r})\nabla\times \textbf{A}]_i
      \ [ \varepsilon^{(2)}({\bf r})\nabla\times \textbf{A}]_j
\end{eqnarray}

Any consistent quantization  of the electromagnetic field should be
able to give the Maxwell equations as the equations of motion of the
electromagnetic field in the Heisenberg picture. It can be  shown
that the classical Maxwell's equation (\ref{a.95}) can be reobtained
in Heisenberg picture if one uses the Hamiltonian (\ref{a21}) and
applies  the commutation relations (\ref{a20}). The Heisenberg
equation for the vector potential gives (see the Appendix B)
\begin{equation}\label{a22}
        \varepsilon^{(1)} {\bf \dot{A}}={\bf \Pi}+\varepsilon^{(1)}\left[
         (\varepsilon^{(1)})^{-1}\ \varepsilon^{(2)}\nabla\times {\bf
        A}\right]^\bot
\end{equation}
which coincides to the definition of the canonical momentum
density in (\ref{a19}). Also the Heisenberg equation for the
conjugate variable $ {\bf \Pi}$ leads to (see the Appendix B)
\begin{eqnarray}\label{a23}
        && {\bf \dot{\Pi}}=\nabla\times\left( \mu^{(1)}({\bf r})\ {\bf
          \dot{A}}({\bf r},t)\right)-\nabla\times\left( \mu^{(2)}({\bf
          r})\nabla\times{\bf A}({\bf r},t)\right)+\nabla\times\left(
       \mu^{(1)}({\bf r}) \nabla\varphi ({\bf r},t)\right)\nonumber\\
&&
\end{eqnarray}
where the Onsager's relations (\ref{a.2}) has been used. Combination
of  the two Heisenberg equations (\ref{a22}) and (\ref{a23}) yields
\begin{eqnarray}\label{a24}
       &&\varepsilon^{(1)} {\bf \ddot{A}}-\varepsilon^{(1)}\left[
       (\varepsilon^{(1)})^{-1}\ \varepsilon^{(2)}\nabla\times  {\bf
       \ddot{A}}\right]^\bot\nonumber\\
       &&=\nabla\times\left( \mu^{(1)}({\bf r})\ {\bf \dot{A}}({\bf
       r},t)\right)-\nabla\times\left( \mu^{(2)}({\bf r})\nabla\times{\bf
       A}({\bf r},t)\right)+\nabla\times\left( \mu^{(1)}({\bf r})
      \nabla\varphi ({\bf r},t)\right)\nonumber\\
&&
\end{eqnarray}
which, regarding the relation (\ref{ap4}) (in Appendix A), is
precisely the Maxwell equation (\ref{a.95}).
\subsection{diagonalization of the Hamiltonian of the system}
To obtain a diagonalized form of the Hamiltonian of the
electromagnetic field, here for the sake of simplicity,  we restrict
ourselves to the special case that the tensors $\varepsilon^{(2)}$
and $\mu^{(1)}$ are identically zero. Then, it is clear from
Eq.(\ref{a5}) that $\varphi=0$ and the Heisenberg equation
(\ref{a24}) is reduced to
\begin{eqnarray}\label{a26}
      -\nabla\times \mu^{(2)}({\bf r})\nabla\times\textbf{A}({\bf
      r},t)=\varepsilon^{(1)}({\bf r}) \ddot{\textbf{A}}({\bf r},t).
\end{eqnarray}
Separating  the time and space coordinates and writing
$\textbf{A}({\bf r},t)=\textbf{\underline{A}}({\bf r}) T(t)$, yields
a differential equation for $T$ as $\ddot{T}=-\omega^2\ T$, where
$\omega^2$ is a proper constant of separation, and the following
equation for $\textbf{\underline{A}}({\bf r})$
\begin{eqnarray}\label{a26.1}
       \nabla\times \mu^{(2)}(\textbf{r})\nabla\times\textbf{\underline{A}}(
       \textbf{r})=\omega^2\varepsilon^{(1)}(\textbf{r})\textbf{\underline{A}}(\textbf{r}),
\end{eqnarray}
In order to solve this equation, we consider a type of eigenvalue
equation as
\begin{equation}\label{a27}
      \nabla\times\mu^{(2)}({\bf r})\nabla\times{\bf F}_k({\bf
      r})=\omega_k^2\ \varepsilon^{(1)}({\bf r})\ {\bf F}_k({\bf r})
\end{equation}
where ${\bf F}_k$ is a squared integrable vector field  that can be
interpreted as an eigenvector field corresponding to an eigenvalue
$\omega_k^2$. The eigenvectors ${\bf F}_k$ and the eigenvalues
$\omega_k^2$ are obtained using the boundary conditions at infinity
and the boundary conditions on the discontinuity surface of the
medium. On the discontinuity surface of the medium, in the absence
of external sorces, the tangential components of the fields $ {\bf
E} , {\bf H}$ and the normal components of the fields ${\bf D}, {\bf
B}$ should be continuous. Using the symmetry  property of the
tensors $ \varepsilon^{(1)}$ and $\mu^{(2)}$, given by (\ref{a.2}),
it can be shown that in the eigenvalue equation(\ref{a27}), those
eigenvectors which are correspond to the different eigenvalues,
satisfy the orthogonality relation
\begin{equation}\label{a28}
\int d^3r\ \varepsilon^{(1)}_{ij}({\bf r})  F_{ki}^*({\bf r})\
 F_{k'j}({\bf r})\sim\delta(\omega_k-\omega_{k'})
\end{equation}
In order to show this, Let us write the complex conjugate of
Eq.(\ref{a27}) for the label  $k'$
\begin{equation}\label{a29}
\nabla\times\mu^{(2)}({\bf r})\nabla\times{\bf F}^*_{k'}({\bf
r})=\omega_{k'}^{* 2}\ \varepsilon^{(1)}({\bf r})\ {\bf
F}^*_{k'}({\bf r}).
\end{equation}
Multiplying Eq.(\ref{a27}) by ${\bf F}^*_{k'}({\bf r})$ and
Eq.(\ref{a29}) by ${\bf F}_{k}({\bf r})$ on the left, then
subtracting and integrating the obtained results on the unbounded
space and  using the boundary conditions at infinity give us
\begin{equation}\label{a30}
(\omega_k^2-\omega^{*2}_{k'})\int d^3r\ \varepsilon^{(1)}_{ij}({\bf
r}) F_{k'i}^*({\bf r})\
 F_{kj}({\bf r})=0.
\end{equation}
For $k=k'$ the integral in the left hand of (\ref{a30}) has a
positive real value. To prove this, let $ C({\bf r})$ be a complex
tensor of the second rank  defined by
\begin{equation}\label{a31}
\varepsilon^{(1)}({\bf r})=C^\dag({\bf r})\ C({\bf r})
\end{equation}
where $C^\dag({\bf r})$ is the Hermitian conjugate of $C({\bf r})$.
Substitution $\varepsilon^{(1)}$ from (\ref{a31}) into the integral
in  (\ref{a30}), one can write
\begin{equation}\label{a32}
\int d^3r\ \varepsilon^{(1)}_{ij}({\bf r}) F_{ki}^*({\bf r})\
 F_{kj}({\bf r})=\int d^3r\ |C({\bf r}){\bf F}_k({\bf r})|^2\ >0
\end{equation}
provided that the tensor $\varepsilon^{(1)}$ and accordingly $C({\bf
r})$ are assumed to be almost everywhere invertible. Equations
(\ref{a30}) and (\ref{a32}) clearly imply the reality of the
eigenvalues $\omega_k^2$ and the
orthogonality relation (\ref{a28}).\\

Although, applying (\ref{a30}) and (\ref{a32}), the orthonormality
  relation
\begin{equation}\label{a33}
\int d^3r\ \varepsilon^{(1)}_{ij}({\bf r})  F_{ki}^*({\bf r})\
 F_{k'j}({\bf r})=\delta_{kk'}
\end{equation}
 can not be proved for  those  eigenvectors which are correspond to
 the
 degenerate eigenvalues,
  in many cases it is possible to construct a complete set of the vector fields
 satisfying the eigenvalue equation (\ref{a27}) and
 the orthonormality  relation (\ref{a33}).
 In fact in those cases that the
tensor $C({\bf r})$, defined by (\ref{a31}), is a hermitian tensor,
that is $C^\dag=C$, it can be shown that the differential operator
\begin{equation}\label{a34}
      \hat{K}= [C({\bf r})]^{-1}\ \nabla\times\mu^{(2)}({\bf r})\
      \nabla\times\ [C({\bf r})]^{-1}
\end{equation}
is a hermitian operator on the Hilbert space
\begin{equation}\label{a35}
       \Omega=\{ {\bf g}:R^3\rightarrow R^3: \hspace{1.00 cm}\int d^3r {\bf
       g}^*({\bf r})\cdot\ {\bf g}({\bf r})<\infty\}
\end{equation}
where the inner product in $\Omega$ is given  by
\begin{equation}\label{a36}
       \forall \ {\bf f}\ ,\ {\bf g}\in \Omega \hspace {1.00 cm}
       \langle{\bf f}|{\bf g}\rangle=\int d^3r \ {\bf f}^*({\bf r})\cdot
       {\bf g}({\bf r})
\end{equation}
Therefore there is   a complete set of vector fields in $\Omega$
that satisfy the eigenvalue equation
\begin{equation}\label{a37}
      [C({\bf r})]^{-1}\ \nabla\times\mu^{(2)}({\bf r})\ \nabla\times\
      [C({\bf r})]^{-1}\ {\bf f}_k=\omega_k^2\ {\bf f}_k
\end{equation}
and the orthonormality relation
\begin{equation}\label{a38}
       \int d^3r\   {\bf f}_{k}^*({\bf r})\cdot
       {\bf f}_{k'}({\bf r})=\delta_{kk'}.
\end{equation}
It is clear from (\ref{a27}) and (\ref{a37}) that the vector field
${\bf F}_k({\bf r})$ will satisfy  Eq.(\ref{a27})  iff   $ C({\bf
r}){\bf F}_k({\bf r})$ satisfies the eigenvalue equation
(\ref{a37}). Accordingly, at least, in those cases that the tensor
$C({\bf r})$ is a hermitian tensor, one can construct a complete set
of the squared integrable vector fields which satisfy the eigenvalue
equation (\ref{a27}) and the orthonormality relation (\ref{a33}).

It should be noted that, as (\ref{a27}) shows, those eigenvector
fields which are correspond to the nonzero eigenvalues satisfy the
gauge condition
\begin{equation}\label{a40}
      \nabla\cdot[\varepsilon^{(1)}({\bf r}){\bf F}_k({\bf r})]=0.
\end{equation}
Regarding  to the completeness of the eigenvector fields one can
expand the projection operator $P^\bot_{ij}({\bf r},{\bf r'})$
,given by (\ref{a11}), as
\begin{equation}\label{a41}
       P^\bot_{ij}({\bf r},{\bf r'})=\sum'_{k}\sum_{l}\ \varepsilon^{(1)}_{jl}({\bf
       r'})\ F^*_{ki}({\bf r})\ F_{kl}({\bf r'})=\sum'_{k}\sum_{l}\
       \varepsilon^{(1)}_{jl}({\bf r'})\ F_{ki}({\bf r})\ F^*_{kl}({\bf
       r'}).
\end{equation}
where $\displaystyle  \sum'_{k}$ denote the summation  over those
eigenvector fields which are correspond to the nonzero eigenvalues
in (\ref{a27}). The condition (\ref{a40}) shows that this  expansion
is compatible with the transversality relations (\ref{a13}) and
(\ref{a14}). Also one can expand the vector potential ${\bf A}$ and
the canonical momentum density ${\bf \Pi}$ as
\begin{eqnarray}\label{a42}
      &&{\bf A}({\bf r},t)= \sum'_k\ \sqrt{\frac{\hbar}{2\omega_k}} \left [
       a_k (0)e^{-\imath\omega_k t}{\bf F}_k({\bf r})+
         a^\dag_k(0) e^{\imath\omega_k t}\ {\bf F}^*_k({\bf r})\right]\nonumber\\
          &&{\bf \Pi}({\bf r},t)=\imath\ \varepsilon^{(1)}({\bf r})\sum'_k\
          \sqrt{\frac{\hbar\omega_k}{2}}\left[ a^\dag_k(0)\ e^{\imath\omega_k
         t}\ {\bf F}^*_k({\bf r})- a_k(0)\ e^{-\imath\omega_k t}\ {\bf
        F}_k({\bf r})\right]\nonumber\\
     &&
\end{eqnarray}
where $a_k$ and $a_k^\dag$ are the annihilation and creation
operators of the bi-anisotropic non-dispersive magnetodielectric
medium. Using the canonical commutation relations (\ref{a20}) and
the expansions (\ref{a41}) and (\ref{a42}), the commutation
relations between $a_k$ and $a_k^\dag$  is  easily obtained as
\begin{equation}\label{a43}
      [a_k(t),\ a_k^\dag(t)]=\delta_{kk'}
\end{equation}
Regarding (\ref{a40}), one can  see that the gauge condition
$\nabla\cdot[ \varepsilon^{(1)}({\bf r}){\bf A}({\bf r},t)]=0$ and $
\nabla\cdot{\bf \Pi}({\bf r},t)=0$ are satisfied by the expansions
(\ref{a42}). Now inserting the expansions (\ref{a42}) in the
Hamiltonian (\ref{a21})  and applying Eqs.(\ref{a27}) and
(\ref{a33}),  we reach to the diagonalized form for  the Hamiltonian
of the system (in the case $\varepsilon^{(2)}=\mu^{(1)}=0$) as the
following
\begin{equation}\label{a44}
      H_F=\sum_k\ \hbar\omega_k\ a^\dag_k a_k
\end{equation}
where the normal ordering has been applied.

\subsection{Homogeneous bulk material}

For a homogeneous  bulk material, that is when the tensors
$\varepsilon^{(1)}$ and $\mu^{(2)}$ are independent of  the
position vector ${\bf r}$, it is easy to show that the
eigenvalues, $\omega_k^2$, in Eq.(\ref{a27}) are the roots of the
following determinant
\begin{equation}\label{a45}
       \textrm{det}\left[ \Lambda({\bf q}, \mu^{(2)})-\omega_k^2
       \,\varepsilon^{(1)}\right]=0,
\end{equation}
where
\begin{equation}\label{a45.1}
       \Lambda_{ij}({\bf
       q},\mu^{(2)})=-\epsilon_{i\alpha\beta}\ \epsilon_{rsj}
       \ \mu^{(2)}_{\beta r}\ q_\alpha\ q_s.
\end{equation}
Let us denote the roots of the determinant in (\ref{a45}) by
$\omega_\rho({\bf q})$, where $\rho$ labels the different roots of
the determinant and ${\bf q}$ is an arbitrary three dimensional
vector. Then, the  normalized eigenvector field corresponds to the
eigenvalue $\omega_\rho({\bf q})$ can be written as
\begin{equation}\label{a46}
      {\bf F}(\rho,\lambda,{\bf q},{\bf r})=\frac{{\bf X}(\rho,\lambda,{\bf
      q})}{\sqrt{{\bf X}^\dag(\rho,\lambda,{\bf q})\varepsilon^{(1)}{\bf
      X}(\rho,\lambda,{\bf q})}} e^{\imath {\bf q}\cdot{\bf r}}
\end{equation}
where ${\bf X}(\rho,\lambda,{\bf q})$ is a three dimensional vector
 satisfying the  algebraic eigenvalue equation
\begin{equation}\label{a47}
      \left[ \Lambda({\bf q},\mu^{(2)})-\omega_\rho^2({\bf q})\
      \varepsilon^{(1)}\right]{\bf X}(\rho,\lambda,{\bf q})=0
\end{equation}
and $\lambda$ indicates the probable  degeneracy of the eigenvalue
$\omega_\rho^2({\bf q})$ which is known as the polarization of the
photon. Here the label $k$ in Eq.(\ref{a27}) is represented by a
triplet $(\rho ,\lambda, {\bf q})$
 and  the orthonormality relation (\ref{a33}) is expressed as
\begin{equation}\label{a48}
       \int d^3r\ \varepsilon^{(1)}_{ij} \  F_{i}^*({\rho ,\lambda,{\bf q}},{\bf
        r})\
        F_{j}({\rho' ,\lambda',{\bf q'}},{\bf r})=\delta_{\rho\rho'}\delta_{\lambda\lambda'}\delta({\bf q}-{\bf
       q'})
\end{equation}
The vector potential and the canonical momentum density ${\bf \Pi}$
can be expanded in terms of the eigenvector fields (\ref{a46}) as
\begin{eqnarray}\label{a49}
       &&{\bf A}({\bf r},t)= \sum'_{\rho} \sum_{\lambda} \int d^3q\
        \sqrt{\frac{\hbar}{2\omega_\rho({\bf q})}} \left [ a(\rho,\lambda,{\bf q})\
         e^{-\imath\omega_\rho({\bf q}) t}\ {\bf F}(\rho ,\lambda, {\bf q},{\bf
          r})+
          \textrm{H.C} \right]\nonumber\\
          &&{\bf \Pi}({\bf r},t)=\imath\ \varepsilon^{(1)}\sum'_{\rho} \sum_{\lambda}\int d^3q\
         \sqrt{\frac{\hbar\omega_\rho({\bf q})}{2}}\left[ a^\dag(\rho,\lambda, {\bf
        q})\  e^{\imath\omega_\rho({\bf q}) t}\  {\bf F}^*({\rho ,\lambda,{\bf
       q}},{\bf r})- \textrm{H.C}\right]\nonumber\\
       &&
\end{eqnarray}
Where $\displaystyle \sum'_\rho$ denotes the summation over the
nonzero roots in (\ref{a45}). Finally, the Hamiltonian of the
electromagnetic field in the presence of the anisotropic homogeneous
magnetodielectric medium can be written as
\begin{equation}\label{a50}
       H=\sum_{\rho,\lambda}\ \int d^3q \ \hbar\omega_\rho({\bf q})\
       a^\dag(\rho,\lambda,
       {\bf q}) a(\rho,\lambda, {\bf q})
\end{equation}

\section{Generalization of the quantization in the presence of external charges}

The quantization method discussed in the previous  sections can be
generalized  in the presence of $N$ point external charged
particles. This  is useful, particularly, when we are concerned with
the spontaneous emission of the atomic systems within the
magnetodielectric media. In the presence of $N$ charged particles
with charges $ q_1 , q_2 , ...,q_N $ and masses $m_1, m_2 , ...,m_N$
the generalization of the Hamiltonian (\ref{a21}) is
\begin{eqnarray}\label{a50.01}
       &&H(t)=\frac{1}{2} \int d^3r\left\{ [ \varepsilon^{(1)}({\bf
        r})]^{-1}_{ij}\ \Pi_i({\bf r},t)\ \Pi_j({\bf r},t)+
         \mu^{(2)}_{ij}({\bf r})\ ( \nabla\times {\bf A})_i\ ( \nabla\times
          {\bf A})_j\right\}\nonumber\\
            &&+\frac{1}{2}\int d^3r\   [ \varepsilon^{(1)}({\bf
             r})]^{-1}_{ij}\left\{\Pi_i({\bf r},t) \left[ \varepsilon^{(2)} ({\bf
              r})\nabla\times {\bf A}\right]_j+ \left[ \varepsilon^{(2)} ({\bf r
             })\nabla\times {\bf A}\right]_i\Pi_j({\bf r},t)\right\}\nonumber\\
            && +\frac{1}{2}  \int d^3r \  [ \varepsilon^{(1)}({\bf
           r})]^{-1}_{ij} \ [\varepsilon^{(2)} ({\bf r})\nabla\times {\bf A}]_i
         \ [ \varepsilon^{(2)}({\bf r})\nabla\times {\bf A}]_j\nonumber\\
        &&+\sum_{i=1}^N\frac{[{\bf p}_i-q_i{\bf A}({\bf x}_i
       ,t)]^2}{2m_i}+\sum_{i=1}^N q_i\ V({\bf x}_i ,t)
\end{eqnarray}
where ${\bf x}_i$ and $ {\bf p}_i$ are the position and momentum
operators of the $i$'th particle that satisfy the usual commutation
relations
\begin{equation}\label{a50.02}
      [{\bf x}_i(t)\ ,\ {\bf p}_j(t)]=i\hbar \delta_{ij}\textrm{I},
\end{equation}
where $\textrm{I}$ is the unit matrix. In this case the commutation
relations between the cartesian components of the conjugate
variables ${\bf A}$ and ${\bf \Pi}$ are the same as (\ref{a20}). In
Hamiltonian (\ref{a50.01}), $V({\bf x}_i ,t)$ is the coulomb
potential produced by the other particles at the place of the $i$'th
particle. This coulomb potential is related to the Green function
introduced by (\ref{a6}) as
\begin{equation}\label{a50.03}
      V({\bf x}_i ,t)= \sum_{j\neq i} q_j\  G({\bf x}_i , {\bf x}_j)
\end{equation}
As the before section,  in the special case that
$\varepsilon^{(2)}=\mu^{(1)}=0$, the conjugate variables ${\bf A}$
and ${\bf \Pi}$ can be expanded in terms of the eigenvectors ${\bf
F}_k$ satisfying Eq.(\ref{a27}) as
\begin{eqnarray}\label{a50.04}
       &&{\bf A}({\bf r},t)= \sum'_k\ \sqrt{\frac{\hbar}{2\omega_k}} \left [
         a_k (t){\bf F}_k({\bf r})+
          a^\dag_k(t) \ {\bf F}^*_k({\bf r})\right]\nonumber\\
          &&{\bf \Pi}({\bf r},t)=i\ \varepsilon^{(1)}({\bf r})\sum'_k\
        \sqrt{\frac{\hbar\omega_k}{2}}\left[ a^\dag_k(t)\ {\bf F}^*_k({\bf
       r})- a_k(t)\ \ {\bf F}_k({\bf r})\right]
\end{eqnarray}
 From the commutation
relations (\ref{a20}) and the expansions  (\ref{a41}), it is clear
that  the annihilation and creation operators $a_k , a^\dag_k$
satisfy  the same commutation rules as (\ref{a43}). Now by inserting
the expansions (\ref{a50.04}) into (\ref{a50.01}) and using the
eigenvalue equation (\ref{a27}), the generalized  Hamiltonian
(\ref{a50.01}) (in the special case
$\varepsilon^{(2)}=\mu^{(1)}=0)$, is reduced to
\begin{equation}\label{a50.05}
      H=\sum_k\ \hbar\omega_k\ a^\dag_k a_k+\sum_{i=1}^N\frac{[{\bf
       p}_i-q_i{\bf A}({\bf x}_i ,t)]^2}{2m_i}+\sum_{i=1}^N q_i\ V({\bf
      x}_i ,t)
\end{equation}
It should be pointed out that, in the Heisenberg picture,  the time
dependence of the operators $ a_k(t) , a^\dag_k(t)$, appeared  in
the expansions (\ref{a50.04}), is no longer sinusoidal as the before
section. The time evolution of $a_k(t)$ and any other dynamical
variable , in the Heisenberg picture, should be obtained by using
the Hamiltonian (\ref{a50.05}) together with the commutation
relations (\ref{a43}) and (\ref{a50.02}).

\section{Spontaneous emission of a two-level atom within an anisotropic magnetodielectric medium}

In this section, using the quantization method discussed  in the
before section, the decay rate of an initially excited two-level
atom in the presence of an anisotropic non-dispersive
magnetodielectric medium, is investigated. In the  Hamiltonian
(\ref{a50.05}), suppose that, except one of the particles, all the
remainder have sufficiently large masses, so that they can be taken
approximately in fixed positions. Then, in the electric dipole
approximation \cite{32,33}, the Hamiltonian (\ref{a50.05}) can be
written as
\begin{equation}\label{a50.1}
       H=\sum_k\ \hbar\omega_k\ a^\dag_k a_k+\frac{{\bf p}^2}{2m}+\phi({\bf
       x})-\frac{e}{m}{\bf p}\cdot{\bf A}({\bf r}_0 )
\end{equation}
where $e$ and $m$ are the charge and mass of the particle that can
move, respectively. $\displaystyle \phi({\bf x})=e \sum_{q_j\neq e}
q_j\ G({\bf x} , {\bf x}_j)$ is the coulomb potential at the place
of the moving particle due to the other fixed particles.  In
(\ref{a50.1}) the position vector ${\bf r}_0$ is the center of the
reign over which the moving particle is free to move. Such a
collection of particles can constitute a one electron atom such that
${\bf r}_0$ points to the center of the atom. In the Hamiltonian
(\ref{a50.1}) the term $\frac{e^2}{m}{\bf A}^2({\bf r}_0)$ has been
ignored , because in the electric dipole approximation this term
dose not affect the decay rate of the atom \cite{33}. To calculate
the spontaneous emission of an atom, we restrict ourselves to the
ideal model of a two-level atom. In this model the basis of the
Hilbert space of the atom is contained two kets denoted by
$|1\rangle$ and $|2\rangle$. The kets $|1\rangle$ and $|2\rangle$
are the eigenstates of the Hamiltonian of the atom corresponding to
the eigenvalues $E_1$ and $E_2$, respectively. For the two-level
atom the Hamiltonian (\ref{a50.1}) can now be rewritten as
\cite{32,33}
\begin{eqnarray}\label{a50.2}
       H&=&H_F+H_{at}+H'\nonumber\\
         H_F&=&\sum_k\ \hbar\omega_k\ a^\dag_k a_k\nonumber\\
          H_{at}&=&\frac{{\bf p}^2}{2m}+\phi({\bf x})= \hbar \omega_0\
         \sigma^\dag \sigma\nonumber\\
        H'&=&[-i \omega_0  \sigma\ {\bf d}\cdot {\bf A}({\bf r}_0) + \textrm{H.C}]
\end{eqnarray}
where $\sigma=|1\rangle\langle2|\ ,\ \sigma^\dag=|2\rangle\langle 1|
$ are pauli operators, $\omega_0=\frac{E_2-E_1}{\hbar}$ is the
eigenfrequency of the atom and ${\bf d}$ is the electric dipole
moment of the atom. Now  by substituting the expansion
(\ref{a50.04}), for ${\bf A}({\bf r}_0)$, into (\ref{a50.2}) the
Hamiltonian in the rotating wave approximation  reduces to \cite{33}
\begin{equation}\label{a50.3}
       H=\sum_k\ \hbar\omega_k\ a^\dag_k a_k+ \hbar \omega_0\ \sigma^\dag
        \sigma -i \omega_0 e \sum_k \sqrt{\frac{\hbar}{2\omega_k}}[\sigma
       a^\dag_k\ {\bf d}\cdot {\bf F}^*_k({\bf r}_0) - \textrm{H.C}]
\end{equation}
The simplest way to estimate the decay rate of an initially excited
two-level atom is the Weisskopf-Wigner approach\cite{33}. In this
approach , in the schr\"{o}dinger picture, the atom-field state at
time $t$ is taken as
\begin{equation}\label{a50.4}
      |\psi(t)\rangle=c(t)|2\rangle|0\rangle+\sum_k\
      M_k(t)|1\rangle|k\rangle
\end{equation}
where $|0\rangle$ is the vacuum state of the electromagnetic field
and the the coefficients $c(t)$ and $ M_k(t)$ satisfy the initial
conditions $c(0)=1$ , $ M_k(0)=0$. Now  applying the Hamiltonian
(\ref{a50.3}) and  substituting $|\psi(t)\rangle$ from
(\ref{a50.4})in the schr\"{o}dinger equation
\begin{equation}\label{a50.5}
      H|\psi(t)\rangle=i\hbar\frac{\partial}{\partial t}|\psi(t)\rangle
\end{equation}
 for sufficiently large times, we have
\begin{equation}\label{a50.6}
      C(t)= e^{(-i \omega_0 -i\delta\omega-\gamma )t}
\end{equation}
where $\delta\omega$ denotes the Lamb shift and $ \gamma$ is the
decay constant of the atom which are given by
\begin{eqnarray}\label{a50.7}
      \delta\omega=\frac{1}{2\hbar} P\sum_k
      \frac{\omega_k}{\omega_0-\omega_k} \ |{\bf d}\cdot {\bf F}_k({\bf
      r}_0)|^2
\end{eqnarray}
\begin{eqnarray}\label{a50.8}
\gamma=\frac{\pi}{2\hbar} \sum_k \omega_k\
\delta(\omega_k-\omega_0)\  |{\bf d}\cdot {\bf F}_k({\bf r}_0)|^2
\end{eqnarray}
where $P$ denotes the Cauchy principle value \cite{25}.

As an example let us assume that the center of the atom is located
at the center of a very small spherical hole of radius R within the
anisotropic  magnetodielectric medium. Taking the origin of
coordinates at the center of the hole, then the tensors
$\varepsilon^{(1)}$ and $ \mu^{(2)}$  in spherical coordinates  have
the form
\begin{equation}\label{a51}
      \varepsilon^{(1)}_{ij}({\bf r})=\left\{ \begin{array}{cc}
                                     \varepsilon_0 &\hspace{1.00 cm} r<R \\
                                     \\
                                     \varepsilon_{ij} & \hspace{1.00 cm}r>R \\
                                   \end{array}\right.
\end{equation}
\begin{equation}\label{a52}
       \mu^{(2)}_{ij}({\bf r})=\left\{ \begin{array}{cc}
                                     \frac{1}{\mu_0} &\hspace{1.00 cm} r<R \\
                                     \\
                                    \mu_{ij} & \hspace{1.00 cm}r>R \\
                                   \end{array}\right.
\end{equation}
where $\varepsilon_0$, $\mu_0$ are the permittivity and permeability
of the vacuum and $\varepsilon_{ij}$ , $\mu_{ij}$ are two constant
tensors which describe the electric and magnetic properties of the
medium around the atom. Applying  these tensors in the eigenvalue
equation (\ref{a27}), the following equation is deduced
\begin{equation}\label{a53}
       \nabla\times \mu\ \nabla\times {\bf F}_k-\omega_k^2\ \varepsilon
       {\bf F}_k=\nabla\times (\mu-\frac{\textrm{I}}{\mu_0})\ \theta(R-r)\
       \nabla\times{\bf F}_k-\omega^2_k\ (\varepsilon-\textrm{I}\varepsilon_0)\
       \theta(R-r)\ {\bf F}_k
\end{equation}
where $\theta(R-r)$ is the unit step function and $I$ is the unit
tensor. If the radius $R$ is assumed to be  very smaller than the
wavelength $\lambda=\frac{2\pi c}{\omega_0}$ then, the following
approximations may be used for the i'the cartesian component of the
right hand of (\ref{a53})
\begin{eqnarray}\label{a54}
       &&\left[-\omega^2_k\ (\varepsilon-I\varepsilon_0)\ \theta(R-r)\ {\bf
        F}_k({\bf r})\right]_i\simeq -\omega^2_k\
         (\varepsilon_{ij}-\delta_{ij}\varepsilon_0)\ \theta(R-r)\ {
          F}_{kj}({0})\nonumber\\
           &&\left[\nabla\times (\mu-\frac{I}{\mu_0})\ \theta(R-r)\
            \nabla\times{\bf
            F}_k\right]_i\nonumber\\
          &&\simeq\epsilon_{i\alpha\beta}\epsilon_{m ns}( \mu_{\beta
         m}-\frac{\delta_{\beta m}}{\mu_0})\left[-
        \frac{x_\alpha}{r}\delta(R-r)\ { F}_{ks,n}(0)+\theta(R-r){
      F}_{ks,\alpha n}(0)\right]\nonumber\\
&&
\end{eqnarray}
where ${ F}_{ks,n}(0)$ and ${F}_{ks,\alpha n}(0)$ denote
$\frac{\partial F_{ks}}{\partial x_n}(0)$ and $\frac{\partial^2
F_{ks}}{\partial x_\alpha\partial x_n}(0)$, respectively. The
solution of the equation (\ref{a53}) can now be written as the sum
of two parts. One part is the solution of the homogeneous equation
\begin{equation}\label{a55}
      \nabla\times \mu\ \nabla\times {\bf F}_k-\omega_k^2\ \varepsilon
      {\bf F}_k=0
\end{equation}
that is
\begin{equation}\label{a55.1}
      {\bf F}_k({\bf r})\equiv {\bf F}(\rho,\lambda,{\bf q},{\bf r})=\frac{{\bf X}(\rho,\lambda,{\bf
       q})}{\sqrt{{\bf X}^\dag(\rho,\lambda,{\bf q})\ \varepsilon\ {\bf
       X}(\rho,\lambda,{\bf q})}} e^{i {\bf q}\cdot{\bf r}}
\end{equation}
 with  ${\bf X}(\rho,\lambda,{\bf q})$
given by (\ref{a45}) and (\ref{a47}), where the tensors
$\varepsilon^{(1)}$ and $ \mu^{(2)}$  should now be  replaced by
$\varepsilon $ and $ \mu$ , respectively. The second part of the
solution (\ref{a53}) is the response to the inhomogeneity term in
the right hand of (\ref{a53}). The cartesian component of this part
can be written as
\begin{equation}\label{a56}
       F_{ki}({\bf r})=\int  d^3r' \tilde{G}_{ij}({\bf r}\ ,\ {\bf r'})\ {J}_j({\bf
       r'})\hspace{1.50 cm} i=1,2,3
\end{equation}
where $J_i$ is the sum of two approximated terms in (\ref{a54}) and
$ \tilde{G}_{ij}({\bf r},{\bf r'})$ is the Green tensor related to
Eq.(\ref{a53}) satisfying
\begin{equation}\label{a57}
      \left[ \epsilon_{i\alpha\beta}\epsilon_{m n s}\ \mu_{\beta
        m}\frac{\partial^2}{\partial x_\alpha \partial x_n}-\omega_k^2 \
        \varepsilon_{is}\ \right]\tilde{G}_{sj}({\bf r}\ ,\ {\bf
      r'})=\delta_{ij}\delta({\bf r}-{\bf r'})
\end{equation}
Using the technique of Fourier transform, the Green tensor
$\tilde{G}$ is easily obtained as
\begin{equation}\label{a58}
\tilde{G}({\bf r}  ,\ {\bf r'})= \lim_{\eta\rightarrow
0^+}\frac{1}{(2\pi)^3}\int_{-\infty}^{+\infty}\ d^3p \frac{1}{[
\Lambda({\bf p},\mu)-\omega^2_\rho({\bf q})\ \varepsilon-i\eta
\textrm{I}]}e^{i {\bf p}\cdot({\bf r}-{\bf r'})}
\end{equation}
where $I$ is the identity tensor and $\Lambda({\bf p},\mu)$ is
defined by (\ref{a45.1}) with $\mu^{(2)}=\mu$. Using (\ref{a54})
 , (\ref{a55.1}) and (\ref{a56}),   the i'th cartesian component of the solution of Eq.(\ref{a53})
 can now be written as
\begin{eqnarray}\label{a59}
       &&{F}_{k i}({\bf r})=\frac{{ X}_i(\rho ,\lambda,{\bf
         q})}{\sqrt{{\bf X}^\dag(\rho,\lambda,{\bf q})\ \varepsilon\ {\bf
          X}(\rho,\lambda,{\bf q})}} e^{i {\bf q}\cdot{\bf
           r}}\nonumber\\
            &&-\omega_\rho^2({\bf q})\int_{-\infty}^{+\infty}\ d^3r' \tilde{G}_{ij}({\bf
            r}, {\bf r'})\
             \theta(R-r')(\varepsilon_{jm}-\varepsilon_0\delta_{jm})\ \ {
              F}_{k m}( 0)\nonumber\\
              &&+\int_{-\infty}^{+\infty}\ d^3r' \tilde{G}_{ij}({\bf r}, {\bf
               r'})\left(\epsilon_{j\alpha\beta}\epsilon_{m ns}( \mu_{\beta
               m}-\frac{\delta_{\beta m}}{\mu_0})\right)\theta(R-r'){
               F}_{k s,\alpha n}( 0)\nonumber\\
              &&+\int_{-\infty}^{+\infty}\ d^3r' \tilde{G}_{ij}({\bf r}, {\bf
             r'})\left(\epsilon_{j\alpha\beta}\epsilon_{m ns}( \mu_{\beta
           m}-\frac{\delta_{\beta m}}{\mu_0})\right)\left[-
         \frac{x'_\alpha}{r'}\delta(R-r')\ {
        F}_{k s,n}( 0)\right]\nonumber\\
     &&
\end{eqnarray}
In (\ref{a58}) and (\ref{a59}) the eigenfrequency $\omega_k$
  has been replaced approximately by $\omega_\rho({\bf q})$, that is, the roots of the determinant
$[ \Lambda({\bf q}, \mu)-\omega_\rho^2({\bf q})\ \varepsilon] $. Now
substituting (\ref{a58}) into (\ref{a59}), after integrating with
respect to the variable ${\bf r'}$, it can be shown
straightforwardly that the third term in (\ref{a59}) is negligible
for sufficiently  small value of $R$ and accordingly  for the i'the
cartesian component of the eigenfunctions ${\bf F}_k({\bf r})$, for
$R\ll 1$, one can write
\begin{eqnarray}\label{a60}
      &&{F}_{k i}( {\bf r})=\frac{{ X}_i(\rho ,\lambda,{\bf
         q})}{\sqrt{{\bf X}^\dag(\rho,\lambda,{\bf q})\ \varepsilon\ {\bf
          X}(\rho,\lambda,{\bf q})}} e^{i {\bf q}\cdot{\bf
           r}}\nonumber\\
          &&+\frac{1}{2\pi^2}\lim_{\eta\rightarrow
           0^+}\left\{\int_{-\infty}^{+\infty}\ d^3p \ \left( e^{i{\bf
            p}\cdot{\bf r}} \ \frac{\sin|{\bf p} R|}{|{\bf p}|^3}\right)[
             \Lambda({\bf p},\mu)-\omega^2_\rho({\bf q})\
             \varepsilon-i\eta I]^{-1}_{ij}\right\}\nonumber\\
             &&\times\left[-\omega_\rho^2({\bf
             q})(\varepsilon_{jm}-\varepsilon_0\delta_{jm})\ \ {
             F}_{k m}( 0)+\left(\epsilon_{j\alpha\beta}\epsilon_{m n
           s}( \mu_{\beta m}-\frac{\delta_{\beta m}}{\mu_0})\right){
         F}_{k s,\alpha n}( 0)\right]\nonumber\\
      &&
\end{eqnarray}
Differentiation with respect to the cartesian coordinates $x_\delta
, x_\gamma$ from the both sides of
 (\ref{a60}) yields
\begin{eqnarray}\label{a61}
       &&{F}_{k i,\delta\gamma}( {\bf r})=\frac{{ X}_i(\rho ,\lambda,{\bf
         q})}{\sqrt{{\bf X}^\dag(\rho,\lambda,{\bf q})\ \varepsilon\ {\bf
          X}(\rho,\lambda,{\bf q})}}(-q_\delta\ q_\gamma) e^{i {\bf q}\cdot{\bf
           r}}\nonumber\\
               &&+\frac{1}{2\pi^2}\lim_{\eta\rightarrow
                0^+}\left\{\int_{-\infty}^{+\infty}\ d^3p \ \left( -p_\delta\
                 p_\gamma \ e^{i{\bf p}\cdot{\bf r}} \ \frac{\sin|{\bf p} R|}{|{\bf
                 p}|^3}\right)[ \Lambda({\bf p},\mu)-\omega^2_\rho({\bf q})\
                  \varepsilon-i\eta I]^{-1}_{ij}\right\}\nonumber\\
                &&\times\left[-\omega_\rho^2({\bf
               q})(\varepsilon_{jm}-\varepsilon_0\delta_{jm})\ \ {
               F}_{k m}( 0)+\left(\epsilon_{j\alpha\beta}\epsilon_{m n
             s}( \mu_{\beta m}-\frac{\delta_{\beta m}}{\mu_0})\right){
           F}_{k s,\alpha n}( 0)\right]\nonumber\\
          &&
\end{eqnarray}
Consistency condition at ${\bf r}=0$ for the relations (\ref{a60})
and (\ref{a61}) gives
\begin{eqnarray}\label{a62}
       &&\Gamma^{(1)}_{im}\ F_{k  m}(0)=\frac{{ X}_i(\rho ,\lambda,{\bf
         q})}{\sqrt{{\bf X}^\dag(\rho,\lambda,{\bf q})\ \varepsilon\ {\bf
          X}(\rho,\lambda,{\bf q})}}+\Delta^{(1)}_{i s\alpha n}\ F_{k s,\alpha
        n}( 0)\nonumber\\
        &&
\end{eqnarray}
and
\begin{eqnarray}\label{a63}
        \Delta^{(2)}_{i \delta \gamma, s \alpha n} F_{k s,\alpha n}( 0)=-\frac{{ X}_i(\rho ,\lambda,{\bf
         q})(q_\delta\ q_\gamma)}{\sqrt{{\bf X}^\dag(\rho,\lambda,{\bf q})\ \varepsilon\ {\bf
          X}(\rho,\lambda,{\bf q})}}+\Gamma^{(2)}_{i \delta \gamma ,m}\
       F_{k m}( 0),
\end{eqnarray}
respectively, where the summation should be done over the repeated
indices and the tensors $\Gamma^{(1)}\ , \Gamma^{(2)}\ ,
\Delta^{(1)}$ and $ \Delta^{(2)}$ are given by
\begin{eqnarray}\label{a64}
       &&\Gamma^{(1)}_{im}=\delta_{i m}+\frac{\omega_\rho^2({\bf
         q})}{2\pi^2}\lim_{\eta\rightarrow 0^+}\left\{\int_{-\infty}^{+\infty}\
          d^3p \ \left(  \ \frac{\sin|{\bf p} R|}{|{\bf p}|^3}\right)[
           \Lambda({\bf p},\mu)-\omega^2_\rho({\bf q})\
          \varepsilon-i\eta I]^{-1}_{ij}\right\}\nonumber\\
       &&\times(\varepsilon_{jm}-\varepsilon_0\delta_{jm})
\end{eqnarray}
\begin{eqnarray}\label{a65}
      &&\Delta^{(1)}_{i s\alpha n}=\frac{1}{2\pi^2}\lim_{\eta\rightarrow
        0^+}\left\{\int_{-\infty}^{+\infty}\ d^3p \ \left(  \ \frac{\sin|{\bf
          p} R|}{|{\bf p}|^3}\right)[ \Lambda({\bf p},\mu)-\omega^2_\rho({\bf
          q})\
          \varepsilon-i\eta \textrm{I}]^{-1}_{ij}\right\}\nonumber\\
         &&\times\left(\epsilon_{j\alpha\beta}\epsilon_{m n s}( \mu_{\beta
        m}-\frac{\delta_{\beta m}}{\mu_0})\right)
\end{eqnarray}
\begin{eqnarray}\label{a66}
      &&\Gamma^{(2)}_{i \delta \gamma ,m}=\frac{\omega_\rho^2({\bf
        q})}{2\pi^2}\lim_{\eta\rightarrow 0^+}\left\{\int_{-\infty}^{+\infty}\
         d^3p \ \left( p_\delta\ p_\gamma \  \ \frac{\sin|{\bf p} R|}{|{\bf
          p}|^3}\right)[ \Lambda({\bf p},\mu)-\omega^2_\rho({\bf q})\
         \varepsilon-i\eta \textrm{I}]^{-1}_{ij}\right\}\nonumber\\
      &&\times(\varepsilon_{jm}-\varepsilon_0\delta_{jm})
\end{eqnarray}
\begin{eqnarray}\label{a67}
      &&\Delta^{(2)}_{i \delta \gamma, s \alpha n}=\delta_{is}
       \delta_{\delta\alpha} \delta_{\gamma
        n}\nonumber\\
         &&+\frac{1}{2\pi^2}\lim_{\eta\rightarrow
          0^+}\left\{\int_{-\infty}^{+\infty}\ d^3p \ \left( p_\delta\
           p_\gamma \  \ \frac{\sin|{\bf p} R|}{|{\bf p}|^3}\right)[
          \Lambda({\bf p},\mu)-\omega^2_\rho({\bf q})\ \varepsilon-i\eta
         \textrm{I}]^{-1}_{ij}\right\}\nonumber\\
        &&\times\left(\epsilon_{j\alpha\beta}\epsilon_{m n s}( \mu_{\beta
       m}-\frac{\delta_{\beta m}}{\mu_0})\right)
\end{eqnarray}
From (\ref{a63}) one can obtain $ F_{k s',\alpha' n'}( 0)$  in terms
of the cartesian component of ${\bf F}( 0)$ as
\begin{eqnarray}\label{a68}
      &&F_{k s',\alpha' n'}( 0)=\nonumber\\
       &&-\frac{1}{\sqrt{{\bf X}^\dag(\rho,\lambda,{\bf q})\ \varepsilon\
        {\bf X}(\rho,\lambda,{\bf q})}}\left[ (\Delta^{(2)})^{-1}_{s'\alpha' n', j
          \delta \gamma}{ X}_j(\rho,\lambda ,{\bf q})(q_\delta\
          q_\gamma)\right]\nonumber\\
         &&+\left[(\Delta^{(2)})^{-1}_{s'\alpha' n',j \delta
        \gamma}\Gamma^{(2)}_{j  \delta \gamma ,m}\ F_{k m}(
       0)\right]
\end{eqnarray}
where the tensor $(\Delta^{(2)})^{-1}_{s'\alpha' n',i \delta
\gamma}$ is defined by
\begin{eqnarray}\label{a69}
      (\Delta^{(2)})^{-1}_{s'\alpha' n',j \delta \gamma}\ \Delta^{(2)}_{j
        \delta \gamma, s \alpha
      n}=\delta_{\alpha\alpha'}\delta_{ss'}\delta_{nn'}
\end{eqnarray}
Now, combination of (\ref{a62}) and (\ref{a68}) yields
\begin{eqnarray}\label{a70}
      &&Q_{im}F_{k m}( 0)=\frac{{ X}_i(\rho,\lambda, {\bf
       q})}{\sqrt{{\bf X}^\dag(\rho,\lambda,{\bf q})\ \varepsilon\ {\bf
        X}(\rho,\lambda,{\bf q})}}\nonumber\\
         &&-\frac{1}{\sqrt{{\bf X}^\dag(\rho,\lambda,{\bf q})\ \varepsilon\ {\bf
         X}(\rho,\lambda,{\bf q})}}\left[ \Delta^{(1)}_{i s \alpha n}\
        (\Delta^{(2)})^{-1}_{s \alpha  n ,j \delta \gamma}\ {
        X}_j(\rho,\lambda,
      {\bf q}) \ q_\delta\ q_\gamma\right]
\end{eqnarray}
where
\begin{eqnarray}\label{a71}
      Q_{im}=\Gamma^{(1)}_{im}- \Delta^{(1)}_{i s \alpha n}\
        (\Delta^{(2)})^{-1}_{s \alpha  n ,j \delta \gamma}\ \Gamma^{(2)}_{j
      \delta \gamma , m}
\end{eqnarray}
Finally, using the relation (\ref{a70}), one can write the $m$'th
cartesian component of $F(\rho,\lambda,{\bf q} , 0)$ as follows
\begin{eqnarray}\label{a72}
      &&F_{k m}( 0)\equiv F_m(\rho , \lambda , {\bf q},0)=\frac{1}{\sqrt{{\bf X}^\dag(\rho,\lambda,{\bf q})\
        \varepsilon\ {\bf X}(\rho,\lambda,({\bf q})}}\nonumber\\
        &&\times \left[ Q^{(-1)}_{mi}\ {
        X}_i(\rho,\lambda,
         {\bf q})
         - Q^{(-1)}_{m i}\Delta^{(1)}_{i s \alpha n}\
        (\Delta^{(2)})^{-1}_{s \alpha  n ,j \delta \gamma}\ {
        X}_j(\rho,\lambda,
      {\bf q}) \ q_\delta\ q_\gamma\right]
\end{eqnarray}
Consequently, Applying the final result (\ref{a72}) in
(\ref{a50.8}), the decay rate of an initially excited two-level
atom located at the center of a very small spherical cavity within
an anisotropic  homogeneous magnetodielectric medium is estimated
as
\begin{eqnarray}\label{a73}
      \gamma=\frac{\pi}{2\hbar} \sum_{\rho,\lambda} \int _{-\infty}^{+\infty} d^3q \
       \omega_\rho({\bf q})\ \delta(\omega_\rho({\bf q})-\omega_0)\ |{\bf
      d}\cdot {\bf F}(\rho ,\lambda, {\bf q} , 0|^2
\end{eqnarray}
It can be shown, straightforwardly,  in the case of an isotropic
dielectric medium the result ({\ref{a73}) coincides to the
spontaneous emission of the atom computed by Glauber and et al in
the limit $R\rightarrow 0$\cite{25}.
\section{summary}
Proposing an appropriate Lagrangian density, a canonical
quantization of the electromagnetic field in the presence of a
non-dispersive bi-anisotropic inhomogeneous magnetodielectric medium
was introduced. The quantization was achieved  for  both in the
absence and in the presence of the atomic systems. The spontaneous
emission of a two-level atom embedded in a bi-anisotropic
homogeneous magnetodielectric medium was investigated. It was argued
that, the form of Maxwell's equations in the curved space-time  in
the absence of a responsive medium , when the cartesian coordinates
is used, are similar to the form of Maxwell's equations in the flat
space-time in the presence of a bi-anisotropic magnetodielectric
medium. Therefore the quantization scheme discussed in this paper is
applicable in the curved space-time  in the absence of responsive
media. This is useful particularly to investigate the effect of the
curvature of the space-time on the quantum properties of the
electromagnetic field.
\renewcommand{\theequation}{A\arabic{equation}}
\setcounter{equation}{0}
\section*{Appendix A : Derivation of the Hamiltonian (\ref{a21})}

By definition, the Hamiltonian of the electromagnetic field  is
written as
\begin{equation}\label{ap1}
       H(t)=\int d^3r [ {\bf \Pi}({\bf r},t)\cdot{\bf \dot{A}}({\bf
       r},t)]-L(t)
\end{equation}
where the Lagrangian $L(t)$ is given by (\ref{a18}). If, using
(\ref{a19}), the time derivative of the vector potential   is
obtained in terms of the dynamical variables ${\bf \Pi}$ and ${\bf
A}$ , then (\ref{ap1}) reads to
\begin{eqnarray}\label{ap2}
      &&H(t)=\frac{1}{2} \int d^3r \left[ (\varepsilon^{(1)})^{-1}_{ij}
       \Pi_i\ \Pi_j + \mu^{(2)}_{ij}\ (
        \nabla\times {\bf A})_i\ ( \nabla\times {\bf A})_j\right]\nonumber\\
         &&+\frac{1}{2} \int d^3r\left\{ {\bf \Pi}\cdot
           \left[(\varepsilon^{(1)})^{-1}\ \varepsilon^{(2)}\nabla\times{\bf
            A}\right]^\bot+\ \ \left[(\varepsilon^{(1)})^{-1}\
             \varepsilon^{(2)}\nabla\times{\bf A}\right]^\bot\cdot
              {\bf \Pi}\right\}\nonumber\\
               &&+\frac{1}{2} \int d^3r \
               \varepsilon^{(1)}_{ij} \left[(\varepsilon^{(1)})^{-1}\
               \varepsilon^{(2)}\nabla\times{\bf A}\right]^\bot_i\
              \left[(\varepsilon^{(1)})^{-1}\
             \varepsilon^{(2)}\nabla\times{\bf A}\right]^\bot_j\nonumber\\
            &&+\frac{1}{2}\int d^3r \int d^3r' \ G({\bf r}, {\bf r'})\
           \nabla\cdot\left[\varepsilon^{(2)}({\bf r})\nabla\times {\bf A}({\bf
          r},t)\right]\ \left[\nabla'\cdot\varepsilon^{(2)}({\bf
         r'})\nabla\times {\bf A}({\bf r'},t)\right]\nonumber\\
       &&
\end{eqnarray}
Because ${\bf \Pi}$ is purely transverse, the second term  is
clearly equivalent to
\begin{eqnarray}\label{ap3}
        \frac{1}{2} \int d^3r\left\{ {\bf \Pi}\cdot
         \left[(\varepsilon^{(1)})^{-1}\ \varepsilon^{(2)}\nabla\times{\bf
        A}\right]+\ \ \left[(\varepsilon^{(1)})^{-1}\
      \varepsilon^{(2)}\nabla\times{\bf A}\right]\cdot {\bf \Pi}\right\}
\end{eqnarray}
According to the definition (\ref{a9}) and the constraint
(\ref{a5}) one can write
\begin{eqnarray}\label{ap4}
      \nabla\varphi=\left[(\varepsilon^{(1)})^{-1}\
      \varepsilon^{(2)}\nabla\times\ {\bf A}\right]^\|
\end{eqnarray}
and therefore from (\ref{a10}) the third term in (\ref{ap2}) is
equal to
\begin{eqnarray}\label{ap4.1}
      &&\frac{1}{2} \int d^3r \ \varepsilon^{(1)}_{ij}
       \left[(\varepsilon^{(1)})^{-1}\ \varepsilon^{(2)}\nabla\times{\bf
        A}-\nabla\varphi\right]_i\ \left[(\varepsilon^{(1)})^{-1}\
        \varepsilon^{(2)}\nabla\times{\bf
       A}-\nabla\varphi\right]_j\nonumber\\
      &&
\end{eqnarray}
which, after integrating by parts, is reduced to
\begin{eqnarray}\label{ap5}
       &&\frac{1}{2}\int d^3r [\varepsilon^{(1)}]^{-1}_{ij}\
         [\varepsilon^{(2)}\nabla\times{\bf A}]_i\
          [\varepsilon^{(2)}\nabla\times{\bf A}]_j+\frac{1}{2}\int d^3r \
         \varphi\  \nabla\cdot[ \varepsilon^{(2)}\nabla\times{\bf A}
        ]\nonumber\\
       &&
\end{eqnarray}
Now applying Eq. (\ref{a5}) the second term in (\ref{ap5}) cancels
the last term in the Hamiltonian (\ref{ap2}) and one can reach to
the form given by (\ref{a21}).
\renewcommand{\theequation}{B\arabic{equation}}
\setcounter{equation}{0}
\section*{Appendix B: Derivation of the Heisenberg equation (\ref{a22}) }

According to the Hamiltonian (\ref{a21}) the time evolution of the
$i$'th cartesian component of the vector potential contains two
parts. The first part is due to the commutator of $A_i$ and the
first term of the Hamiltonian (\ref{a21}). Using the symmetry
feature of the tensor $\varepsilon^{(1)}$ and the commutation
relations (\ref{a20}), this part can be expressed as
\begin{eqnarray}\label{bp1}
      &&\frac{i}{\hbar}\int d^3r' \ [ \varepsilon^{(1)}({\bf
        r'})]^{-1}_{\alpha\beta} \ \Pi_\alpha({\bf r'},t)\
         \left[\Pi_\beta({\bf
         r'},t)\ ,\ A_i({\bf r},t)\right]\nonumber\\
        &&=\int d^3r' \ [ \varepsilon^{(1)}({\bf r'})]^{-1}_{\alpha\beta}\
        P^\bot_{i\alpha}({\bf r}, {\bf r'})\ \Pi_\beta({\bf
       r'},t)=\left[(\varepsilon^{(1)}({\bf r}))^{-1}\ \Pi({\bf
      r},t)\right]_i
\end{eqnarray}
where the fact that  the vector field $ (\varepsilon^{(1)})^{-1}
{\bf \Pi}$ is purely $\varepsilon^{1}$-transverse field has been
used in the last step. The second part of the time evolution of
$A_i$ is caused by the commutator of the third and fourth terms of
the Hamiltonian (\ref{a21}) and $A_i$, which using the symmetry
property of the tensor $\varepsilon^{(1)}$ can be written as the
following
\begin{eqnarray}\label{bp2}
      &&\frac{i}{\hbar}\int d^3r' \ [ \varepsilon^{(1)}({\bf
       r'})]^{-1}_{\alpha\beta}\left\{ [\Pi_\alpha({\bf r'},t)\ ,\
        A_i({\bf r},t)]\left( \varepsilon^{(2)}({\bf r'})\nabla'\times{\bf
         A}({\bf
          r'},t)\right)_\beta\right\} \nonumber\\
          &&=\int d^3r' \ \ [ \varepsilon^{(1)}({\bf r'})]^{-1}_{\alpha\beta}\
          P^\bot_{i\alpha}({\bf r}, {\bf r'})\  \left( \varepsilon^{(2)}({\bf
          r'})\nabla'\times{\bf A}({\bf r'},t)\right)_\beta\nonumber\\
        &&=\left[ (\varepsilon^{(1)})^{-1}\  \varepsilon^{(2)}\ \nabla\times
       {\bf A}\right]^\bot_i
\end{eqnarray}
where the definition  (\ref{a12.1}) has been applied in the last
step. Now by adding (\ref{bp1}) and (\ref{bp2}) the Heisenberg
equation (\ref{a22}) is deduced.
\renewcommand{\theequation}{C\arabic{equation}}
\setcounter{equation}{0}
\section*{Appendix C: Derivation of the Heisenberg equation (\ref{a23})}

The Hamiltonian (\ref{a21}) shows that the time derivative of
 the $i$'th cartesian component of ${\bf
\Pi}$ contains three parts. Regarding the symmetry property $
\mu^{(2)}_{\alpha\beta}=\mu^{(2)}_{\beta\alpha}$ and the commutation
relations (\ref{a20}), the first part is
\begin{eqnarray}\label{cp1}
       &&\frac{i}{2\hbar}\int d^3r'\ \mu^{(2)}_{\alpha\beta}({\bf
        r'})\left[ \left( \nabla'\times{\bf A}({\bf r'} ,t)\right)_\alpha\
         \left( \nabla'\times{\bf A}({\bf r'} ,t)\right)_\beta \ , \
          \Pi_i({\bf r},t)\right]\nonumber\\
           &&=\frac{i}{\hbar}\int d^3r'\ \mu^{(2)}_{\alpha\beta}({\bf r'})
            \left( \nabla'\times{\bf A}({\bf r'} ,t)\right)_\alpha\ \left[
            \left( \nabla'\times{\bf A}({\bf r'} ,t)\right)_\beta \ , \
           \Pi_i({\bf r},t)\right]\nonumber\\
          &&=-\int d^3r'\ \mu^{(2)}_{\alpha\beta}({\bf r'})\left(
         \nabla'\times{\bf A}({\bf r'} ,t)\right)_\alpha\ \epsilon_{\beta p
       q}\ \partial'_p P^\bot_{qi} ({\bf
      r'},{\bf r})\nonumber\\
\end{eqnarray}
Integrating by parts and applying (\ref{a15}), the final result in
(\ref{cp1})  is reduced to
\begin{eqnarray}\label{cp2}
      &&=-\int d^3r'\ P^\bot_{qi} ({\bf r'},{\bf r})\ \left[
       \nabla'\times\mu^{(2)}({\bf r'})\ \nabla'\times {\bf A}({\bf
        r'},t)\right]_q\nonumber\\
       &&=-\left[ \nabla\times\mu^{(2)}({\bf r})\ \nabla\times {\bf A}({\bf
      r'},t)\right]_i
\end{eqnarray}
Another part of  $\dot{\Pi}_i$ is related to the commutator of the
third and forth terms of the Hamiltonian (\ref{a21}) and $\Pi_i$.
Using the symmetry relations $
[\varepsilon^{(1)}]^{-1}_{\alpha\beta}=[\varepsilon^{(1)}]^{-1}_{\beta\alpha}$
and commutation relations (\ref{a20}), this part can be written as
\begin{eqnarray}\label{cp3}
       &&=\frac{i}{\hbar}\int d^3r'\  [ \varepsilon^{(1)}({\bf
         r'})]^{-1}_{\alpha\beta}\ \ \Pi_\alpha({\bf r'},t)\ [\
          \left(\varepsilon^{(2)}({\bf r'})\nabla'\times{\bf A}({\bf
           r'},t)\right)_\beta\ ,\ \Pi_i({\bf r},t)\ ]\nonumber\\
           &&=-\ \int d^3r'\  [ \varepsilon^{(1)}({\bf
            r'})]^{-1}_{\alpha\beta}\ \varepsilon^{(2)}_{\beta s}({\bf r'})\
             \Pi_\alpha({\bf r'},t)\ \epsilon_{spq}\ \partial'_p\
            P^\bot_{qi}({\bf r'}, {\bf r})\nonumber\\
           &&=\ \int d^3r'\  [ \varepsilon^{(1)}({\bf
           r'})]^{-1}_{\alpha\beta}\ \mu^{(1)}_{s \beta}({\bf r'})\
         \Pi_\alpha({\bf r'},t)\ \epsilon_{spq}\ \partial'_p\
       P^\bot_{qi}({\bf r'}, {\bf r})\nonumber\\
     &&
\end{eqnarray}
where the Onsager's  relation
$\varepsilon^{(2)}_{\alpha\beta}=-\mu^{(1)}_{\beta\alpha}$ has been
used. Integration  by parts and using the relations
(\ref{a15}),(\ref{a22}),(\ref{a10}) and (\ref{ap4}) the final result
 in (\ref{cp3}) is  equivalent  to
\begin{eqnarray}\label{cp4}
      &&=\int d^3r'\ P^\bot_{qi}({\bf r'},{\bf r})\ \left[
       \nabla'\times\mu^{(1)}({\bf r'}) (\varepsilon^{(1)}({\bf r'}))^{-1}\
        {\bf \Pi}({\bf r',t})\right]_q\nonumber\\
         &&=\left[ \nabla\times\mu^{(1)}({\bf r}) (\varepsilon^{(1)}({\bf
          r}))^{-1}\ {\bf \Pi}({\bf r,t})\right]_i\nonumber\\
          &&=[\nabla\times\mu^{(1)}({\bf r}){\bf \dot{A}}({\bf r},t)]_i+[
           \nabla\times\mu^{(1)}({\bf r})\nabla\varphi({\bf
           r},t)]_i\nonumber\\
          &&-\left[\nabla\times\mu^{(1)}({\bf r})(\varepsilon^{(1)}({\bf
         r}))^{-1} \varepsilon^{(2)}({\bf r})\
       \nabla\times {\bf A}({\bf r},t)\right]_i\nonumber\\
     &&
\end{eqnarray}
Finally,  the third part of time derivative  of $ \Pi_i$ is caused
by the commutator of the last term of the Hamiltonian (\ref{a21})
and $\Pi_i$ which can be computed as
\begin{eqnarray}\label{cp5}
       &&=\frac{i}{\hbar}\int d^3r'\  [ \varepsilon^{(1)}({\bf
        r'})]^{-1}_{\alpha\beta}\left\{[\ \left(\varepsilon^{(2)}({\bf
         r'})\nabla'\times{\bf A}({\bf r'},t)\right)_\alpha\ ,\ \Pi_i({\bf
          r},t)\ ]\ \left(\varepsilon^{(2)}({\bf r'})\nabla'\times{\bf
           A}({\bf
           r'},t)\right)_\beta\right\}\nonumber\\
           &&=-\int d^3r'\  [ \varepsilon^{(1)}({\bf
           r'})]^{-1}_{\alpha\beta}\ \ \varepsilon^{(2)}_{\alpha s}({\bf r'})
          \ \epsilon_{spq}\
         \partial'_p P^\bot_{qi}({\bf r'}, {\bf r})\
        \left(\varepsilon^{(2)}({\bf r'})\nabla'\times{\bf A}({\bf
      r'},t)\right)_\beta
\end{eqnarray}
Using the onsager's relation
$\varepsilon^{(2)}_{\alpha\beta}=-\mu^{(1)}_{\beta\alpha}$,
integrating by parts and then, applying the relation (\ref{a15}),
this is equal  to
\begin{eqnarray}\label{cp6}
      &&=\int d^3r'\ P^\bot_{qi}({\bf r'}, {\bf r}) \left[
        \nabla'\times\mu^{(1)}({\bf r'}) (\varepsilon^{(1)}({\bf r'}))^{-1}\
         \varepsilon^{(2)}({\bf r'})\ \nabla'\times{\bf A}({\bf
         r'},t)\right]_q\nonumber\\
        && =\left[\nabla\times\mu^{(1)}({\bf r}) (\varepsilon^{(1)}({\bf
       r}))^{-1}\ \varepsilon^{(2)}({\bf r})\ \nabla'\times{\bf A}({\bf
     r},t)\right]_i
\end{eqnarray}
which cancels the last term in (\ref{cp4}). Therefore, adding the
parts (\ref{cp2})), (\ref{cp4}) and (\ref{cp6}) gives  the
Heisenberg equation (\ref{a23}).

\end{document}